\documentstyle[12pt,procsla,psfig]{article}
  
  \catcode`\@=11
  \long\def\@makefntext#1{
  \protect\noindent \hbox to 3.2pt {\hskip-.9pt  
  $^{{\ninerm\@thefnmark}}$\hfil}#1\hfill}		
  
  \def\@makefnmark{\hbox to 0pt{$^{\@thefnmark}$\hss}}  
  	
  \def\ps@myheadings{\let\@mkboth\@gobbletwo
  \def\@oddhead{\hbox{}
  \rightmark\hfil\ninerm\thepage}   
  \def\@oddfoot{}\def\@evenhead{\ninerm\thepage\hfil
  \leftmark\hbox{}}\def\@evenfoot{}
  \def\sectionmark##1{}\def\subsectionmark##1{}}
  
\begin{document}
  
  \centerline{\normalsize\bf TOWARD RADIO DETECTION OF PEV NEUTRINOS ON}
  \baselineskip=16pt
  \centerline{\normalsize\bf THE CUBIC KILOMETER SCALE
  \footnote{Conf. Proc. 7$^{\rm th}$ International Symposium on
  Neutrino Telescopes, Feb. 27 - Mar. 1, 1996.}}

  \vfill
  \vspace*{0.6cm}
  \centerline{\footnotesize GEORGE M. FRICHTER}
  \baselineskip=13pt
  \centerline{\footnotesize\it Bartol Research Institute, University 
  of Delaware}
  \baselineskip=12pt
  \centerline{\footnotesize\it Newark, DE 19716, USA}
  \centerline{\footnotesize E-mail: frichter@lepton.bartol.udel.edu}
  \vspace*{0.3cm}
  \centerline{\footnotesize JOHN P. RALSTON  and  DOUGLAS W. MCKAY}
  \baselineskip=13pt
  \centerline{\footnotesize\it Department of Physics and Astronomy, 
  University of Kansas}
  \centerline{\footnotesize\it Lawrence, KS 66045-2151, USA}
  \centerline{\footnotesize E-mail: ralston@kuphsx.phsx.ukans.edu}
  \centerline{\footnotesize E-mail: mckay@kuphsx.phsx.ukans.edu}

  \vspace*{0.9cm}
  \abstracts{Interactions of ultrahigh energy neutrinos of cosmological 
  origin in large volumes of radio-transparent South Polar ice can be 
  detected via coherent $\check{\rm C}$erenkov emission from accompanying
  electromagnetic showers. A pilot experiment employing buried radio
  receivers has been successfully deployed at the South Pole and data are
  now being collected. The physics of coherent radio emission
  together with the properties of radio-pulse propagation in Antarctic
  ice clearly distinguishes the radio method from
  phototube detection. In the context of the proposed ${\rm km}^3$-scale
  neutrino telescope, these two detection modes provide
  {\it complementary} information on UHE neutrino interactions.}

  \normalsize\baselineskip=15pt
  \setcounter{footnote}{0}
  \renewcommand{\thefootnote}{\alph{footnote}}
  \section{Introduction}
  The case has been made at this conference and elsewhere that the natural
  scale for ultra-high energy (UHE) neutrino astronomy
  is 1 ${\rm km}^3$~\cite{KM3,Gaisser}. 
  Several groups are making significant progress toward the
  goal of ``KM3'' detection by employing smaller scale phototube arrays 
  deep in lake or ocean water and Antarctic ice~\cite{Gaisser}. In
  the context of the overall KM3 effort, these efforts can be regarded as a 
  proving grounds for detection strategies and associated technologies best 
  suited to the task of ``instrumenting a mountain''.

  One promising detection strategy involves observing coherent radio
  $\check{\rm C}$erenkov emission~\cite{Zhelez}. The coherent effect is
  produced for frequencies up to the microwave region (few GHz)~\cite{ZHS}
  and results in a detection efficiency that scales with energy 
  very differently compared with conventional phototube
  detection~\cite{FMR,Price}. 
  Cold Antarctic ice is also remarkably transparent to 
  electromagnetic (EM) radiation at frequencies below a few GHz, having 
  attenuation lengths easily exceeding 1 kilometer~\cite{Atten}, whereas
  absorption at optical frequencies is about one order of magnitude less
  favorable. Scattering lengths for radio and optical frequencies also 
  differ significantly, with virtually no volume scattering of long wavelength
  radiation expected. These factors lead one to consider seriously the
  potential complementarity of the two detection modes in the context of a
  KM3 detector located at the South Pole. In addition to the obvious advantage
  of independently observing events with detectors obeying different physics, 
  unique kinematic infomation about simultaneously observed events
  is available which cannot be achieved by employing either detection mode
  separately.

  As a step toward understanding the operation of radio detectors
  in the Antarctic, a pilot experiment has been deployed this season in
  two of the four AMANDA-B deep bore holes at the South Pole. In section 2,
  we discuss the elements of this deployment and describe its current status.

  Section 3 is devoted to a discussion of the basic scaling
  relations for radio detection, and section 4 compares the overall
  efficiency of radio receivers and phototubes for detecting neutrino
  induced cascades. Important differences between the two methods and
  some new ideas for exploiting those differences in the context of a
  hybrid radio-phototube array are explored in section 5.

  Much excitement in the field of UHE neutrino astronomy
  has been generated recently by the possibility that neutrinos from
  active galactic nuclei (AGN) may dominate the spectrum at energies
  above a few hundred TeV. Section 5 gives a detailed
  estimate for the neutrino event rate seen by a radio receiver buried
  deep in the Antarctic ice cap based on a theoretical model for the
  diffuse AGN flux.

  \section{RICE - A Pilot Radio Experiment$^\dagger$}

\begin{figure}
  \vspace*{20pt}
  \centering
  \mbox{\psfig{figure=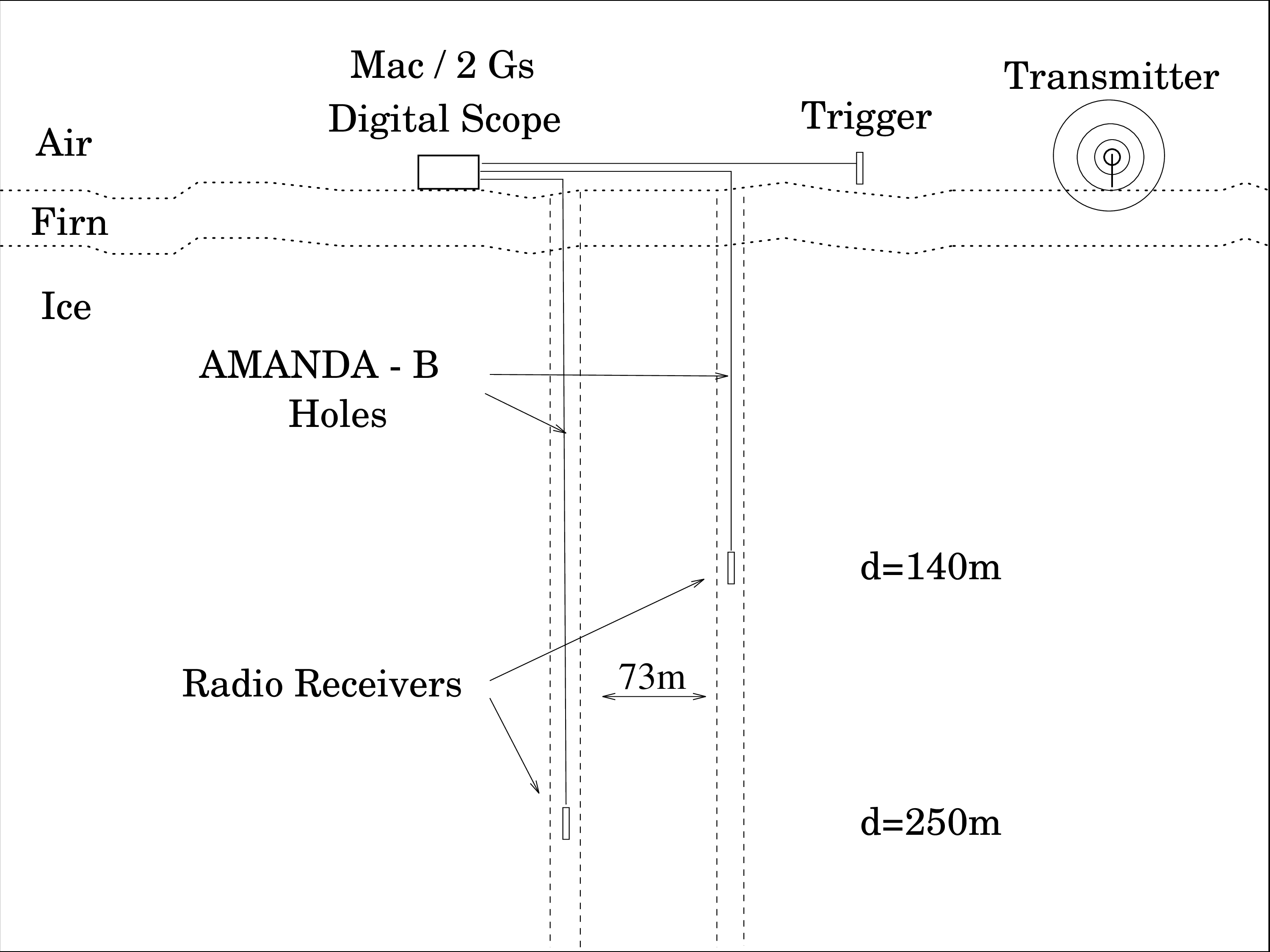,width=10.0cm}}
  \vspace*{0.2truein}           
  \fcaption{1995-96 RICE deployment: consists of three radio receivers
  (two deep, one on surface),
  one pulse transmitter (mobile), and a data acquisition 
  system on the surface. Receivers are connected to the DAQ by low-loss
  transmission lines.}
  \label{fig:sketch}
  \end{figure}

  Figure ~\ref{fig:sketch} indicates the main elements of the 
  RICE 1995-96 deployment.
  The experiment consists of two deep radio receivers at depths
  of 250 meters and 150 meters, a surface mounted ``trigger'' receiver,
  a mobile transmitter, and finally a data acqusition system (DAQ) consisting
  of a 2 Gigasample digital scope driven by a microcomputer. The receivers
  are connected to the DAQ by low loss coaxial transmission lines. Typical 
  data samples collected on each channel are voltages sampled at
  2 GHz over a 60 $\mu{\rm s}$ time window, giving a very detailed
  record of received pulse shapes. 

  The radio receivers were
  designed with simplicity and survivability in mind. Each consists of
  a simple copper cylindrical dipole ``tuned'' such that the expected
  $\lambda/2$ resonance in ice (refractive index = 1.78) is about 130 MHz
  with a 3 dB bandwidth of 10 to 15 MHz. Each dipole is encased in a 
  PVC/epoxy housing for good structural strength and low drag during
  deployment. Amplification of the signal to
  a level suitable for the trip up to the surface occurs by two low
  noise amplification stages (power gain: 36 dB each, 72 dB total) 
  housed inside a steel pressure vessel.
  Power for each amplifier along with the amplified signal are carried on 
  a single coax transmission line.

  The mobile transmitter is capable of producing pulses with a center frequency
  of 120 MHz with a switchable bandwidth of 10 to 50 MHz. Given the limited
  bandwith of the receivers, the shortest pulses will be useful for
  observing the ``impulse response'' of the receivers {\it in situ}. 
  This is a key element in understanding the test results.

  The scientific and technical goals of the current deployment can
  be summarized as follows.

  \begin{itemize}
  \item design and deploy radio receivers suited to the harsh Antarctic
        conditions - electronics must be water tight up to 1 km depth and
        able to survive over-pressures which occur during refreeze,
        as well as temperatures in the $-50^o$ C range.
  \item observe coincident pulses on three data channels and
        characterize the response of the receivers using the transmitter.
  \item identify and characterize sources of ambient radio frequency noise in
        the south polar ice.
  \end{itemize}
 
  At the time of this conference,
  some two to three weeks after deployment of the two deep receivers,
  preliminary data have been collected and we can say that the first of these
  three goals has been achieved. The receivers have survived deployment and
  appear to be operating normally. At this very early stage, some key elements
  of the test remain to be deployed (the surface transmitter and the surface
  receiver), and we eagerly await the results of the full test in the near
  future.

  \section{Basic Scaling Relations for Radio Detection}

  The efficiency for radio detection of neutrino induced EM cascades in ice
  scales very differently with energy compared to phototube detection.
  This fact can be understood by recalling the Frank-Tamm formula for
  $\check{\rm C}$erenkov power radiated per frequency per path length
  by a charge $z$, moving with speed 
  $v=\beta c$, through a medium with refractive index $n$:

  \begin{equation}
  \frac{d^2W}{d\nu dl}=(\frac{4 \pi^2 \hbar}{c}\alpha_{\rm em}) z^2 \nu
  [1-\frac{1}{\beta^2 n^2}]\:.
  \label{eq1}
  \end{equation}
  We immediately see that the available signal power scales with
  frequency and the {\it square} of the charge involved. 
  When considering the power radiated by all the charges in a large EM
  cascade, the high and low frequency limits display different behavior
  due to simple coherence effects.
  For wavelengths small compared with the spatial charge distribution, 
  the phases of arriving $\check{\rm C}$erenkov wavelets from each particle 
  are randomly
  distributed resulting in an incoherent sum of each contribution
  to the total observed power. In this case, 
  the signal power scales with the total number of charged particles,
  which is given roughly by the shower energy $E_s$
  measured in GeV. At the opposite extreme, when wavelengths are
  large in comparison to the shower dimensions, wavelets arriving from
  charged particles are in phase, leaving a coherent
  sum of impulses from the showers' net charge. It is well
  known that EM showers in dense media develop a significant excess of
  negative charge mainly by Compton scattering of shower photons on atomic
  electrons (Bhabha scattering, M{\o}ller scattering, as well as e$^+$+e$^-$
  annihilation also contribute to a lesser degree)~\cite{ZHS,Allan}. 
  The detailed
  Monte Carlo simulation for EM cascades in ice developed by Zas, Halzen and
  Stanev (ZHS)~\cite{ZHS} reveals that
  the net charge developed by a large shower scales with energy like
  $z \sim .3 E_s/{\rm GeV}$. Equation (\ref{eq1}) then says
  that the available signal power
  in the coherent regime is increasing as $E_s^2$. Because of this difference
  in scaling behavior ($E_s$ for photo-detection compared with $E_s^2$ for
  radio detection), at sufficiently large shower energies the power in the
  long-wavelength components of the $\check{\rm C}$erenkov signal will dominate.

  At a distance $R$ from the cascade the signal intensity varies with
  energy as $E_s^2/R^2$ due to spherical spreading. For a fixed detector
  threshold, we see that the maximum detection distance is proportional to
  the energy, $R_{\rm max}\sim E_s$. The detection volume per detector then
  increases with the shower energy {\it cubed}, $V_{\rm det}\sim E_s^3$.
  This is the naive expectation in the absence of signal absorption in
  the medium and geometric effects. Fortunately, absorption lengths for 
  radio frequency pulses in ice at measured Antarctic temperatures
  are known to excede 1 kilometer~\cite{Atten},
  so that the basic $E_s^3$ scaling
  should hold up to volumes per radio receiver of order 1 km$^3$.

  \section{Quantitative Efficiency Estimates}
  \label{secQ}

  In a recent paper we performed detailed simulations
  based on a particular radio antenna and receiver
  which include all factors relevant to estimating detection 
  efficiency and event rates (based on theoretical projections 
  for diffuse neutrino fluxes)~\cite{FMR}.
  One important ingredient of those estimates is a parameterization of
  the signal spectrum (Fourier components of the time dependent electric field)
  given by the ZHS Monte Carlo~\cite{ZHS},

  \begin{figure}
  \centering
  \mbox{\psfig{figure=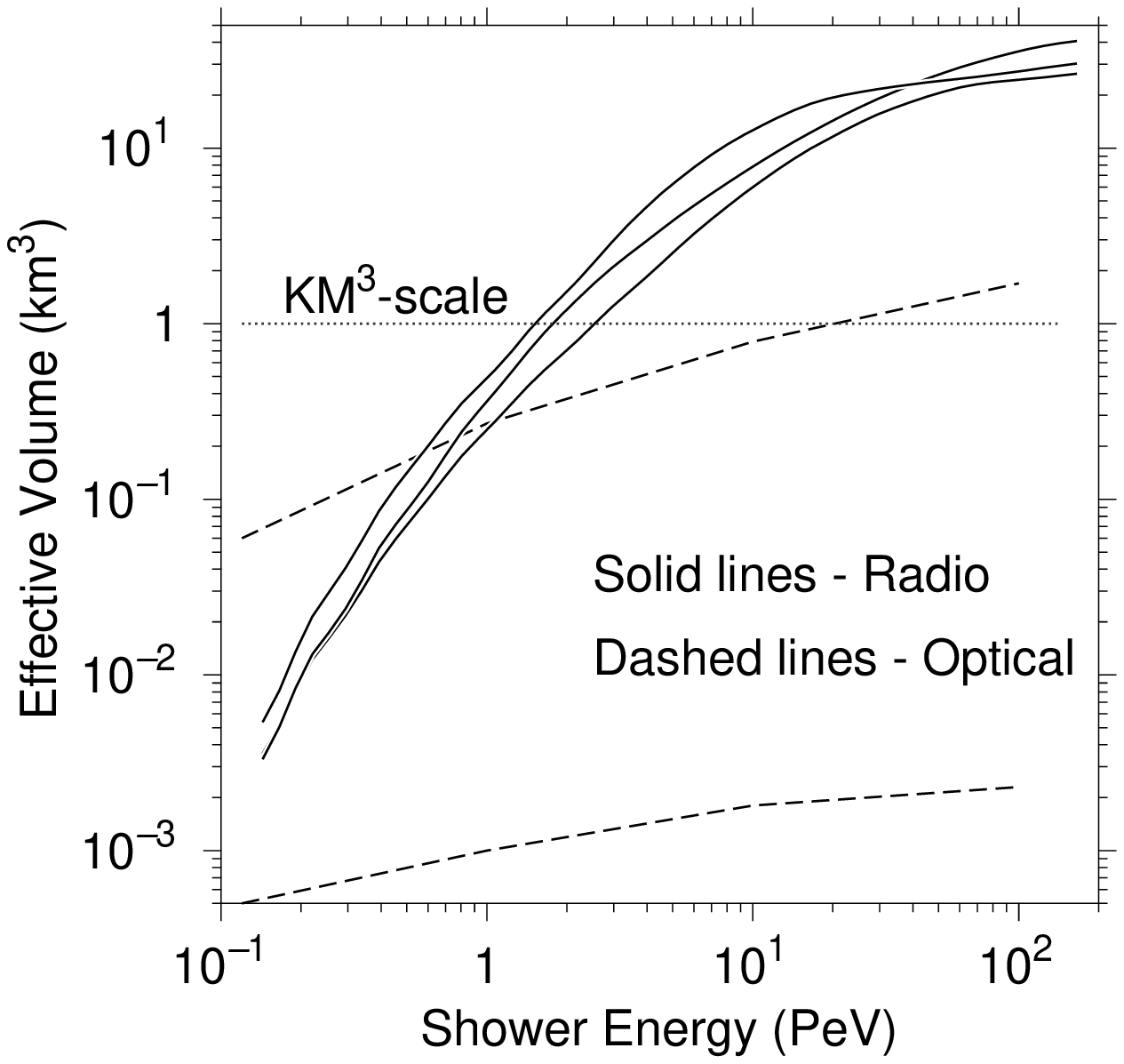,width=10.0cm,clip=}}
  \fcaption{Comparison of detection efficiencies of radio antennas versus
  phototubes. Optical estimates are from Ref. 6}
  \label{fig:volume}
  \end{figure}

  \begin{equation}
  R|\vec{E}(\nu,R,\theta_c)|=\frac{0.55\times 10^{-7}(\nu/\nu_o)}
  {1+0.4(\nu/\nu_o)^2}\frac{E_s}{1{\rm TeV}} {\rm exp}[-\frac{1}{2}
  (\frac{\theta-\theta_c}{\theta_o}\frac{\nu}{\nu_o})^2]
  (\frac{{\rm V}}{{\rm MHz}})\:,
  \label{eqSpec}
  \end{equation}
  where $R$ is the distance between observer and cascade,
  $\nu_o=500$ MHz, $\theta_o\sim 2.4^o$, $\theta_c=56^o$ is the
  $\check{\rm C}$erenkov angle in ice, and $\theta$ is the angle between 
  the shower axis and the observer. The target volume surveyed by a
  single radio receiver is an important measure of efficiency.
  By reciprocity, one can regard this also as the volume occupied by a 
  detectable signal following a neutrino event. As the radio pulse propagates
  outward from the cascade site, it sweeps out a `fat' conical volume,
  $V_{\rm signal}$, set by
  the $\check{\rm C}$erenkov angle, the frequency-dependent
  angular half-width, $\Delta\theta=\theta_o\nu_o/\nu$, and the maximum
  range at which the pulse remains above threshold: 

  \begin{equation}
  V_{\rm signal}=\frac{4}{3}\pi R^3_{\rm max} {\rm sin}\theta_c 
  \Delta\theta\:.
  \label{eqVsig}
  \end{equation}
  Choosing a typical frequency of 250 MHz gives a signal volume of 1 km$^3$
  when $R_{\rm max}\sim 1.4$ km. The appropriate value for $R_{\rm max}$
  depends on signal attenuation in the medium as well as the details of
  detector design.

  Figure \ref{fig:volume} compares effective detection volumes for a simple
  biconical antenna (`tuned' to 250 MHz) with a calculation of the same
  quantity for a 3 inch phototube~\cite{FMR,Price}. 
  The three solid curves show the radio
  efficiency for incident neutrino angles (with respect to vertical downward)
  of 120, 140 and 160 degrees (see figure 14 of reference~\cite{FMR}). The
  lower and upper dased curves are for photo-detection with and without
  ice bubbles~\cite{Price}. In the case of ice with bubbles, the radio
  and optical effective volumes are comparable near 100 TeV. A reasonably
  optimistic calculation of the case of ice with no bubbles shows comparable
  effective volumes near 800 TeV. In either event, radio becomes more efficient
  with increasing energy, exceeding either estimate of the phototube effective 
  volume by at least a factor of 10 at 10 PeV.

  \section{Antennas and Phototubes: A Match Made In Heaven?}

  The roughly comparable volumes shown in Fig. \ref{fig:volume} 
  suggest that radio receivers and phototubes 
  might be employed together
  as independent detection modes for a KM3 observatory, offering
  complementary information about neutrino events. At energies
  below 10-100 TeV, where phototube sensitivity dominates, the hybrid KM3
  array would operate by observing mainly
  $\nu_\mu$ induced muons passing through the arrayed volume. Above about 
  100 TeV the radio and optical modules will begin to
  survey increasingly large target masses surrounding the fiducial array
  volume (see Fig. \ref{fig:volume}) for neutrino induced cascades. Until now,
  the advantages of employing a hybrid radio-optical array have not been
  fully appreciated. In this section we hope to point out just a few
  of the interesting possibilities, focusing on electron neutrino events.

  \subsection{Background Rejection}

  One consequence of the poor radio efficiency for $E_s < 100$ TeV
  is that, compared with phototube detection, radio is relatively insensitive 
  to unwanted backgrounds
  from local cosmic ray interactions (atmospheric neutrinos and muons)
  which have steeply falling spectra. In fact, our recent estimate indicates
  less than one background event per year from atmospheric
  neutrinos~\cite{FMR}. Radio-phototube coincidence
  introduces automatic energy discrimination that could be used to reject 
  unwanted physics backgrounds.

  \subsection{Scattering, Event Geometry and Pulse Timing}

  The ability to ``point'' a neutrino telescope amounts to having reliable
  information on the geometry of observed events. This is clearly of
  fundamental importance to UHE neutrino astronomy.
  The `fat' cone geometry described in section 4 together with
  pulse timing are key to 
  obtaining good pointing accuracy from cascade observations. 
  How this information is
  retained for radio versus optical frequency signals differs significantly
  due to the vast difference in the wavelengths involved 
  (a factor of $10^6$).

  Having nicely macroscopic wavelengths (order meters),
  radio waves are expected to suffer no significant
  volume scattering as they propagate over kilometer distances.
  Marcroscopic density variations should
  occur only in the uppermost 100 meters (firn layer) of ice, and will
  affect only a small fraction of potential event geometries. The net
  result is that radio pulses can retain good directional and timing
  information about the event even after propagating over large distances 
  from interaction site to receiver.

  In contrast, scattering opportunities for optical photons appear to
  be plentiful. Scattering in bubble-free ice at the South Pole is 
  expected to be dominated by dust particles with a scattering length
  of about 20 meters (considerably less in dust bands) and a mean
  scattering angle of $<{\rm cos}(\theta)>\sim 0.9$~\cite{Price}. 
  The optical signal begins life with the sharp geometry of the
  original $\check{\rm C}$erenkov cone, but these photons then begin to
  diffuse, spreading out arrival times and eventually becoming isotropic 
  on the KM3 scale. This clearly impacts pulse timing and geometry
  determination. However, the diffusing signal
  leads to improved detection thresholds because more phototubes 
  per event will be `hit' compared with a non-diffusing scenario.

  The complementarity of combining optical and radio signals may lead
  to very interesting enhancements and deserves to be investigated in detail.
  For example, timing information supplied by the radio pulse may
  assist with photodetection, both by allowing thresholds to be adjusted
  advantageously and by knowing when to look in the photodetector output.
  It seems possible that distant events whose diffuse signal would be
  spread out too much in time for straight photodetection might be
  registered by using the extra information from the radio signal.
  More study of this, and attention to the substantial difference between
  electron and muon neutrino events is certainly needed.

  \begin{figure}
  \vspace*{15pt}
  \centering
  \mbox{\psfig{figure=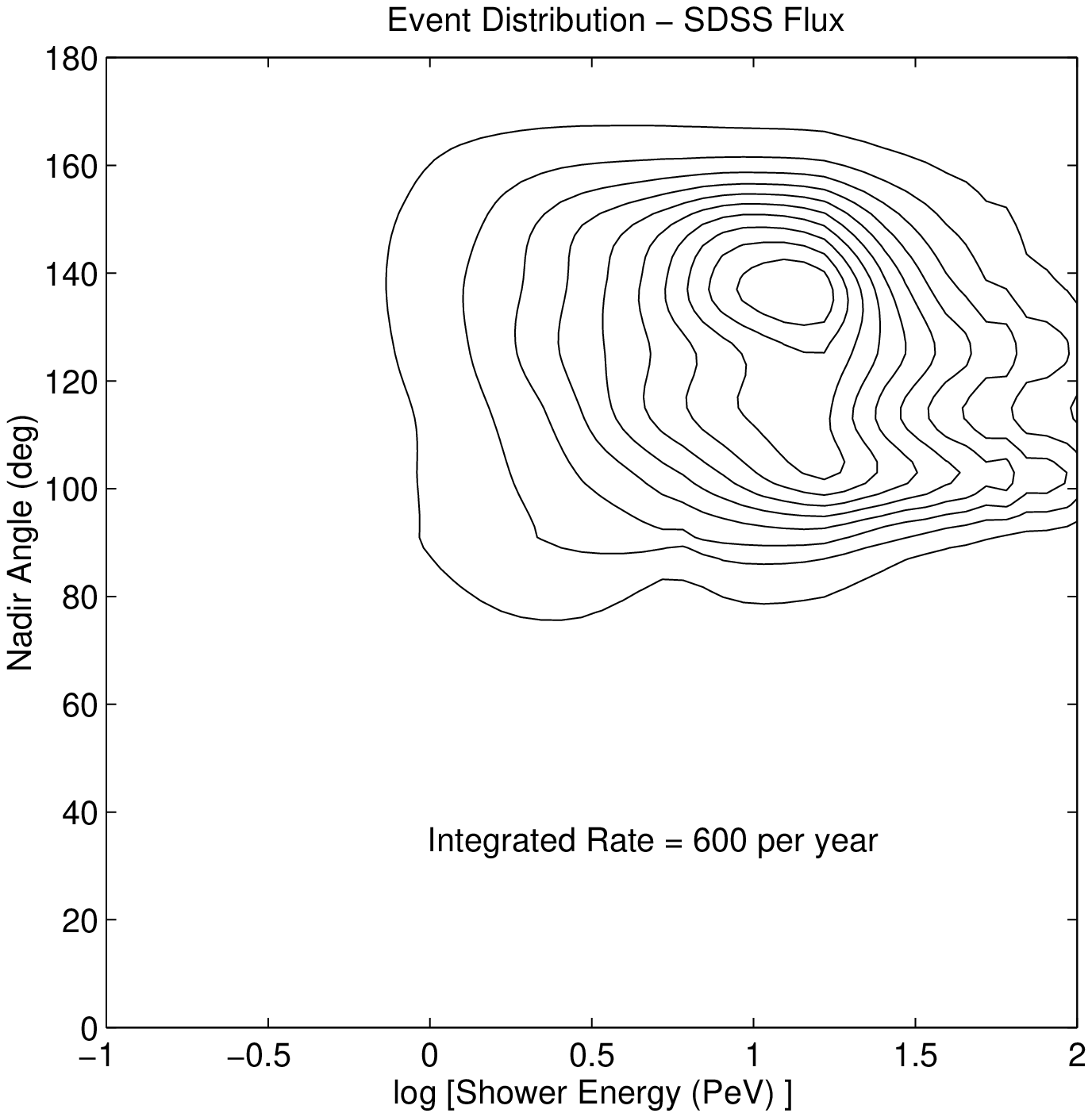,width=8.0cm}}
  \vspace*{0.2truein}           
  \fcaption{The distribution of AGN neutrino events
  as detected by a single antenna located 500 meters below the ice surface
  using the SDSS flux prediction. 
  We plot 10 equally spaced contours of the quantity $dN/d{\rm
  cos}\theta/d{\rm log}E$, with the ratio between the innermost and
  outermost contours equal to 10.}
  \label{fig:rate}
  \end{figure}

  An extremely interesting idea~\cite{frich} involves exploiting
  the different propagation speeds of radio and optical signals in ice.
  The refractive index for radio signals in ice is about 1.8 and about
  1.3 for visible light. If there is no scattering (or if ``leading''
  optical photons are detected~\cite{Mock}), then a signal 100
  meters away from a co-located optical/radio detector arrives a
  ``huge'' 150 ns later in radio than in visible light. At 1 km event
  distance the timing difference becomes 1.5 $\mu$s ! If the timing
  resolution of the detectors is 10 ns, then the two arrival times
  determine the distance to the event vertex to within
  $\pm$ `a few meters'.
  Evaluation of the usefulness of this phenomenon is,
  however, complicated by scattering and
  the resulting broadening of the optical component. 
  We are investigating the use of these techniques not only to enhance
  detection, but also in conjunction with pulse height information to
  more accurately determine event energy~\cite{frich,Mock}. The multiplicity of
  new ``handles'' from joining optical and radio detection is very exciting
  indeed.

  \section{Radio Event Rates - Diffuse AGN Prediction}

  Here we will briefly present some results for radio event rates based on
  models for the diffuse neutrino flux predicted for AGN; details
  can be found elsewhere~\cite{FMR}.
  Figure (\ref{fig:rate}) shows our result based on the
  model of Stecker {\it et al.}~\cite{Steck}
  (SDSS) for a radio receiver buried 500
  meters below the ice surface. The contours indicate constant event
  probabilities as a function of the shower energy (horizontal axis)
  and the incident angle for the incoming neutrino (vertical axis). 
  Events are predicted
  to be clustered at energies between 1 and 50 PeV and angles between
  horizontal and about 60$^o$ above horizontal. Note that almost no
  neutrinos come from below the horizon due to absorption in the Earth,
  except for a few events $< 20^o$ below the horizon at energies below
  10 PeV. Although the angular distribution is skewed by the effects of
  absorption, the increased cross section at UHE yields an overall increase
  in total rate compared to earlier estimates~\cite{FRM}.
  The total rate for the SDSS model is about 600 events per year per detector.
  An array of a few hundred radio detectors in a cubic kilometer volume
  would create a genuine radio neutrino telescope.

  \section{Summary}

  \begin{itemize}
  \item A pilot experiment employing buried radio receivers in the Antarctic 
        was successfully deployed at the South Pole this year.
  \item Differences between radio and phototube detection of UHE cascades
        result both from physics at the signal source and signal
        propagation in the medium. These differences present a unique 
        opportunity to obtain {\it independent} infomation on event 
        kinematics, improving the accuracy of geometry and energy
        estimation and extending the sensitivity of the KM3
        observatories to higher energies.
  \item Diffuse neutrinos from AGN may produce large numbers of events
        in a radio array.
  \end{itemize}

  A unique opportunity exists for KM3 neutrino astronomy in the Antarctic.
  In polar ice we are indeed fortunate to have two complementary and comparably 
  efficient detection modes available to observe neutrino induced cascades.
  The advantages of employing radio and phototube elements on the KM3-scale
  have only begun to be investigated. We
  believe that maximizing the `cost per unit science' function will
  make a compelling case for hybrid radio-phototube arrays.

  \section{Acknowledgements}
  This work was supported in part under Department of Energy Grant Numbers
  DE-FGO2-85-ER 40214 and DE FG02 91 ER 40626.A007,
  and by the Kansas Institute for Theoretical and Computational Science.

  \vspace*{15pt}
  $^\dagger$ List of RICE Collaborators:

  \begin{itemize}
    \item[*] Bartol Research Institute, Newark, DE 19716 USA
    \begin{itemize}
      \item George Frichter, Tim Miller, Lucio Piccirillo, David Seckel, 
            Michele Limon
    \end{itemize}

    \item[*] University of Kansas, Lawrence, KS 66045-2151, USA
    \begin{itemize}
       \item Dave Besson, Alice Bean, Chris Allen, Sergei Kotov, 
             Ilya Kravchenko, Suruj Seunarine
    \end{itemize}
  \end{itemize}

  \end{document}